%% file: paper.tex
\documentclass[conference]{IEEEtran}
\IEEEoverridecommandlockouts

\usepackage{cite}
\usepackage{amsmath,amssymb,amsfonts}
\usepackage{algorithmic}
\usepackage{graphicx}
\usepackage{textcomp}
\usepackage{xcolor}
\def\BibTeX{{\rm B\kern-.05em{\sc i\kern-.025em b}\kern-.08em
    T\kern-.1667em\lower.7ex\hbox{E}\kern-.125emX}}

\usepackage{macros} 

\titlespacing*{\section} {0pt}{0.2ex}{0.2ex}
\titlespacing*{\subsection} {0pt}{0.2ex}{0.2ex}
\IEEEaftertitletext{\vspace{-5.6ex}}
 
\begin{document}

\title{\LARGE{\simnametitle: A Unified Simulator for \\Heterogeneous and Disaggregated LLM Serving Infrastructure}\\[-1.2ex] \thanks{\vspace{-3ex}*These authors contributed equally to this work.}}

\author{
\IEEEauthorblockN{Jaehong Cho*}
\IEEEauthorblockA{\textit{School of Computing} \\
\textit{KAIST}\\
Daejeon, South Korea \\
\href{mailto:jhcho@casys.kaist.ac.kr}{\textcolor{blue}{jhcho@casys.kaist.ac.kr}}}
\and
\IEEEauthorblockN{Hyunmin Choi*}
\IEEEauthorblockA{\textit{School of Computing} \\
\textit{KAIST}\\
Daejeon, South Korea \\
\href{mailto:hmchoi@casys.kaist.ac.kr}{\textcolor{blue}{hmchoi@casys.kaist.ac.kr}}}
\and
\IEEEauthorblockN{Guseul Heo}
\IEEEauthorblockA{\textit{School of Computing} \\
\textit{KAIST}\\
Daejeon, South Korea \\
\href{mailto:gsheo@casys.kaist.ac.kr}{\textcolor{blue}{gsheo@casys.kaist.ac.kr}}}
\and
\IEEEauthorblockN{Jongse Park}
\IEEEauthorblockA{\textit{School of Computing} \\
\textit{KAIST}\\
Daejeon, South Korea \\
\href{mailto:jspark@casys.kaist.ac.kr}{\textcolor{blue}{jspark@casys.kaist.ac.kr}}}
}

\maketitle

\input{body/abstract}
\input{body/intoduction}

\input{body/back}
\input{body/design}

\input{body/example}

\input{body/method}
\input{body/eval}
\input{body/related}

\input{body/conclusion}

\input{body/acknowledgement}

\bibliographystyle{IEEEtran}
\balance
\bibliography{paper}

\input{artifact/ae.tex}

\end{document}

%% file: body/abstract.tex
\begin{abstract}

Large language model (LLM) serving infrastructures are undergoing a shift toward heterogeneity and disaggregation. Modern deployments increasingly integrate diverse accelerators and near-memory processing technologies, introducing significant hardware heterogeneity, while system software increasingly separates computation, memory, and model components across distributed resources to improve scalability and efficiency. As a result, LLM serving performance is no longer determined by hardware or software choices in isolation, but by their runtime interaction through scheduling, data movement, and interconnect behavior. However, understanding these interactions remains challenging, as existing simulators lack the ability to jointly model heterogeneous hardware and disaggregated serving techniques within a unified, runtime-driven framework.

This paper presents \simname, a unified system-level simulator designed to make runtime-driven hardware–software interactions in heterogeneous and disaggregated LLM serving infrastructures explicit and analyzable. \simname embeds serving decisions and hardware behavior into a single runtime loop, enabling interaction-aware modeling of batching, routing, placement, offloading, memory management, and power consumption. The simulator supports extensible integration of emerging accelerators and memory systems through profile-based modeling, while capturing dynamic serving behavior and system-level effects. We validate \simname against real serving deployments, showing that it reproduces key performance, memory, and power metrics with an average error of 0.95\%, while maintaining practical simulation times of around 10 minutes even for complex system configurations. These results demonstrate that \simname provides a practical bridge between hardware innovation and serving-system design, enabling systematic exploration and co-design for next-generation LLM serving infrastructures. Our simulator is available at \href{https://github.com/casys-kaist/LLMServingSim}{\textcolor{blue}{https://github.com/casys-kaist/LLMServingSim}}.

\end{abstract}

\begin{IEEEkeywords}
Large language model (LLM), Inference serving, Heterogeneous hardware, Simulation infrastructure
\vspace{-0.5ex}
\end{IEEEkeywords}


%% file: body/intoduction.tex
\section{Introduction}

Large language model (LLM) serving infrastructures are becoming increasingly \emph{heterogeneous}, moving beyond the homogeneous, GPU-only deployments in most existing systems. 
While NVIDIA GPUs remain central, modern deployments increasingly integrate diverse hardware, including hyperscalers' accelerators such as Google TPU~\cite{tpuv4} and Amazon Inferentia~\cite{amazon-inferentia}, as well as emerging domain-specific NPUs~\cite{sambanova,cerebras,groq,rebellion,furiosa,hyperaccel}. 
In parallel, memory-centric architectures, including processing-in-memory (PIM) and processing-near-memory (PNM) technologies actively developed by memory vendors such as Samsung~\cite{samsung-pim,samsung-pnm} and SK hynix~\cite{skh-pim,skh-pnm}, further expand the heterogeneity of the serving substrate. 
These trends indicate a shift in which LLM serving performance is shaped by the coordinated use of diverse compute and memory devices, making heterogeneity an inherent characteristic of modern LLM serving infrastructure.
Alongside this shift, \emph{disaggregation} has emerged as a fundamental structural pattern in modern LLM serving systems. 
As deployments scale beyond a single server or a device type, computation, memory, and model components are increasingly separated across distributed resources to improve efficiency and flexibility. 
Modern serving techniques reflect this co-evolution by explicitly disaggregating execution through mechanisms such as prefill–decode separation~\cite{distserve, splitwise}, mixture-of-experts with expert parallelism and offloading~\cite{rajbhandari2022deepspeedmoeadvancingmixtureofexpertsinference, sida-moe, 10.1145/3669940.3707267}, and memory-level disaggregation enabled by prefix caching~\cite{cachedattention, promptcache, cacheblend} and remote memory pools~\cite{lia, instattention, oasis, aide}. 
In practice, disaggregation and heterogeneous resource utilization often arise together as complementary design choices for optimizing performance and efficiency, jointly shaping the dominant architectural direction for scaling LLM serving.

\input{table/sim-compare}

In this landscape, a key challenge for researchers designing new hardware and system architectures is to understand how heterogeneous devices and disaggregated system structures interact in practice. 
Performance, efficiency, and scalability are no longer determined by hardware or software choices in isolation, but by how these choices interact at runtime, including their impact on scheduling decisions, memory behavior, and interconnect contention. 
Gaining such insights directly from deployed systems is costly and often impractical, requiring substantial engineering effort and large-scale infrastructure even to explore high-level design trade-offs. 
Consequently, simulation tools play a critical role in enabling early-stage exploration of hardware–software co-design for LLM serving. 
However, as shown in Table~\ref{tab:sim-comparison}, existing simulators fall short of this need, as they do not adequately model the runtime interactions between heterogeneous hardware and disaggregated serving techniques within a unified framework.
To address this gap, we present LLMServingSim 2.0, a unified simulator designed to make hardware–software interactions in heterogeneous and disaggregated LLM serving systems explicit and analyzable. 
LLMServingSim 2.0 embeds serving decisions and hardware behavior within a single, runtime-driven simulation loop, allowing batching, routing, placement, and offloading decisions to adapt dynamically to system conditions. 
By modeling how these decisions evolve and propagate over time, the simulator enables direct exploration of performance and efficiency behaviors that arise from hardware–software interaction, rather than from static configurations. 
As summarized in Table~\ref{tab:sim-comparison}, LLMServingSim 2.0 brings together capabilities that are only partially supported in prior simulators, establishing a unified framework for interaction-aware evaluation of modern LLM serving infrastructures.
We realize these capabilities through a set of modeling contributions that together form the unified framework:

\begin{itemize}[labelindent=0.3em,nolistsep,leftmargin=1.0em]
  \item \textbf{Interaction-awareness.}
  LLMServingSim~2.0 explicitly models the runtime interactions between serving software decisions and heterogeneous hardware behavior, capturing how batching, routing, placement, caching, and offloading adapt to evolving system state and how their effects propagate over time. By treating these interactions as first-class modeling targets, the simulator enables analysis of performance behaviors that arise from feedback between software policies and hardware conditions.

  \item \textbf{Unified modeling of heterogeneity and disaggregation.}
  The simulator provides a unified representation of heterogeneous accelerators, multi-tier memory systems, and disaggregated serving architectures, enabling coherent evaluation of system behaviors that emerge only from their combined interaction. This unified view allows researchers to study how disaggregated serving techniques behave under different hardware compositions without isolating hardware and software effects.

  \item \textbf{Runtime-driven serving dynamics.}
  By embedding serving decisions into a runtime-driven simulation loop, LLMServingSim~2.0 allows system performance to emerge from dynamic request flows, resource contention, and interconnect effects, rather than from static or a priori configurations. This approach captures temporal effects such as queueing, contention amplification, and phase-dependent behavior that are difficult to observe with static models.

  \item \textbf{Extensibility to emerging hardware.}
  Through profile-based operator modeling, LLMServingSim~2.0 supports low-effort integration of new accelerators and memory technologies, enabling evaluation of future hardware designs without restructuring the serving model. This design facilitates rapid exploration of new hardware–software combinations as LLM serving infrastructures continue to evolve.

  \item \textbf{Power-aware modeling.}
  LLMServingSim~2.0 incorporates power modeling into the simulation framework, enabling joint evaluation of performance and energy-related trade-offs in LLM serving systems. By associating compute, memory, and data-movement activities with power characteristics, the simulator allows researchers to analyze how serving strategies, hardware heterogeneity, and disaggregation choices impact power consumption and energy efficiency alongside performance.
\end{itemize}
We evaluate LLMServingSim 2.0 by validating its accuracy, efficiency, and practical usefulness as a system-level simulation tool for LLM serving infrastructure. Specifically, we compare simulated serving performance against real deployments, showing that LLMServingSim 2.0 reproduces key metrics such as throughput, latency breakdowns (e.g., TTFT and TPOT), memory usage, and power consumption with an average error of 0.95\% across representative workloads and hardware platforms. We further demonstrate that the simulator enables efficient exploration of heterogeneous and disaggregated serving configurations, while maintaining practical simulation times on the order of minutes even for complex system setups, without sacrificing architectural fidelity. Together, these results show that LLMServingSim 2.0 serves as a practical bridge between hardware innovation and serving-system design, enabling systematic co-design and early-stage exploration for the evolving LLM serving infrastructure ecosystem.

%% file: table/sim-compare.tex
\begin{table}[t]
\centering
\caption{Comparison of LLM serving simulators}
\label{tab:sim-comparison}
\resizebox{\linewidth}{!}{
\begin{tabular}{l|ccc|ccc|cccc}
\toprule
\textbf{Simulator} 
& \multicolumn{3}{c|}{\textbf{Dissagg.}} 
& \multicolumn{3}{c|}{\textbf{Parallelism}} 
& \multicolumn{4}{c}{\textbf{Modeling}} \\ 
\cmidrule(lr){2-4} \cmidrule(lr){5-7} \cmidrule(lr){8-11}
 & PD & AF & HT & PP/TP & DP & EP & PA & PC & EO & PM \\ 
\midrule
LLMServingSim~\cite{llmservingsim} & \nocheck & \yescheck & \nocheck & \yescheck & \nocheck & \nocheck & \yescheck & \nocheck & \nocheck & \nocheck\\
Vidur~\cite{vidur} & \nocheck & \nocheck & \nocheck & \yescheck & \yescheck & \nocheck & \semicheck & \nocheck & \nocheck & \nocheck \\
APEX~\cite{apex} & \nocheck & \nocheck & \nocheck & \yescheck & \yescheck & \yescheck & \nocheck & \nocheck & \nocheck & \yescheck \\
TokenSim~\cite{tokensim} & \yescheck & \nocheck & \nocheck & \yescheck & \semicheck & \nocheck & \yescheck & \semicheck & \nocheck & \nocheck \\
\textbf{Ours} & \yescheck & \yescheck & \yescheck & \yescheck & \yescheck & \yescheck & \yescheck & \yescheck & \yescheck & \yescheck \\ 
\bottomrule
\end{tabular}
}
\vspace{0.1em}

{\footnotesize
\raggedright
PD: Prefill/Decode Disaggregation, 
AF: Attention/FFN Disaggregation, 

HT: Heterogeneous System,
PP/TP: Pipeline/Tensor Parallelism, 

DP: Data Parallelism, 
EP: Expert Parallelism, 
PA: PagedAttention, 

PC: Prefix Caching, 
EO: Expert Offloading,
PM: Power Modeling.

\raggedright
\yescheck: fully supported,
\nocheck: not supported,
\semicheck: limited or partial support. \par
}
\vspace{-0.1em}
\end{table}

%% file: body/back.tex
\section{Background}

\subsection{LLM Inference Serving: Workflow and System Structure}

Modern LLM inference follows a two-phase execution workflow consisting of a compute-intensive \textit{prefill} stage and a memory-intensive \textit{decode} stage~\cite{attention}.  
Prefill stage processes the input sequence using dense matrix multiplications, while the decode stage generates tokens autoregressively with repeated key-value (KV) cache accesses.
As such, prefill stresses compute throughput, whereas decode is primarily constrained by memory bandwidth, capacity, and KV cache locality.

In practical deployments, data-center scale serving systems concurrently operate tens to hundreds of model instances~\cite{alpaserve, deepserve, spotserve, aegaeon}.
Incoming requests induce dynamic batching and queuing, which result in time-varying compute and memory utilization.  
To meet latency and throughput requirements under such variability, modern serving frameworks employ diverse techniques, including tensor, pipeline, and data parallelism~\cite{megatron-lm}, expert parallelism for mixture of experts (MoE) models~\cite{rajbhandari2022deepspeedmoeadvancingmixtureofexpertsinference,pre-gated-moe,10.1145/3669940.3707267}, prefix caching for KV cache reuse~\cite{cacheblend, impress,kvcachecache, sglang,liu2025lmcacheefficientkvcache,mooncake}, and prefill-decode (PD) disaggregation across separate clusters~\cite{distserve,jin2024pdserveservingdisaggregatedlarge,splitwise,mooncake}.
These techniques reshape execution order, memory access patterns, communication behavior, and device utilization during inference serving.

\subsection{Heterogeneous Accelerators and Memory Hierarchies}

LLM serving today is built on increasingly heterogeneous hardware platforms.  
A single cluster may host GPUs of different generations~\cite{splitwise}, NPUs and TPUs~\cite{tpuv4, dfx,mcbp, hybe} specialized for matrix operations, memory-centric accelerators such as Processing-in-Memory (PIM)~\cite{neupims,attacc,duplex,papi}, and systems augmented with Compute Express Link (CXL)-based devices~\cite{cxl-gpu, directcxl, systematic-cxl, oasis}.  
These accelerators differ substantially in compute capability, on-device memory bandwidth, memory capacity, interconnect throughput, and power characteristics, resulting in a highly non-uniform compute substrate.

Memory hierarchy plays a central role in LLM inference performance.  
KV cache accesses span multiple memory tiers, including accelerator HBM, host DRAM, storage devices, and large CXL memory pools, each with distinct latency and bandwidth characteristics~\cite{impress, mooncake, infinigen, flexgen}.
During decode, the efficiency of KV cache placement, reuse, and migration directly influences latency and throughput.  
For MoE models, expert weights may be partitioned across devices or offloaded to memory-rich units, further shaping the pattern of operator execution and inter-device communication~\cite{pre-gated-moe,10.1145/3669940.3707267,rajbhandari2022deepspeedmoeadvancingmixtureofexpertsinference}.

Together, these trends illustrate that LLM serving operates over a diverse and multi-tiered compute-memory stack.  
Execution behavior, data movement, and device heterogeneity interact in complex ways, motivating the need for modeling tools that capture the characteristics of modern LLM serving.

\section{Motivation}

\subsection{Modeling Complex Runtime Dynamics in LLM Serving}

LLM serving performance is shaped not by isolated operator execution but by the evolution of runtime dynamics driven by incoming requests.
As requests arrive, queues form, batch sizes fluctuate, and schedulers adapt placement decisions based on device and memory pressure.  
KV cache blocks are allocated, reused, or migrated across tiers; prefix caching introduces hit and miss patterns; and MoE models incur token-dependent expert routing.
These tightly coupled events determine time-to-first-token (TTFT), time-per-output-token (TPOT), end-to-end latency distribution, and throughput.  
Static or operator-only performance models cannot capture such temporal effects, including shifting batch composition, locality-dependent KV cache behavior, heterogeneous-device cooperation, or routing imbalance.  
Accurate evaluation requires a simulator capable of representing these interactions as they unfold over time.

\subsection{Limitations of Existing LLM Serving Simulators}

Existing LLM simulation tools fall into two broad categories, each modeling different aspects of the serving stack while leaving key serving behaviors insufficiently captured.
Hardware-centric simulators, such as LLMCompass~\cite{llmcompass} and ADOR~\cite{ador}, enable detailed modeling of low-level accelerator behavior and communication.
However, their focus on static execution prevents them from capturing dynamic serving-time behaviors, such as request arrivals, batching evolution, and KV cache reuse, which are critical to end-to-end serving performance.
System-level simulators, including Vidur~\cite{vidur} and APEX~\cite{apex}, facilitate exploration of optimal execution configurations and scheduling policies at the serving layer. However, their modeling of memory management and heterogeneous deployments is limited, particularly for features that form the core of recent serving systems such as prefix caching and PD disaggregation.
TokenSim~\cite{tokensim} takes a step toward modeling serving dynamics, but its simplified memory abstraction restricts fidelity when studying KV placement, bandwidth contention, and migration.
Finally, LLMServingSim~\cite{llmservingsim} focuses on collecting runtime statistics in a single-instance setting, but it is not designed to study multi-instance or heterogeneous setups, nor serving features such as MoE behavior, prefix caching, or PD disaggregation.


\subsection{Need for a Unified Simulator}

Modern LLM serving systems operate at the intersection of heterogeneous accelerators, multi-tier memory, multiple parallelism schemes (including MoE), prefix caching, and multi-instance routing with disaggregation.  
These components interact in nontrivial ways, producing behaviors that cannot be captured by partial or layer-specific models.  
Supporting hardware-software co-design, policy evaluation, and architectural exploration therefore requires a unified simulator that integrates hardware fidelity, memory hierarchy dynamics, serving techniques, and fully dynamic request flows.  
This need motivates the development of \simname, which provides a cohesive, extensible framework for modeling end-to-end LLM serving behavior on heterogeneous systems.

%% file: body/design.tex
\section{\simname}
\label{sec:llmservingsim2.0}

\begin{figure}[]
  \centering
  \includegraphics[width=0.95\linewidth]{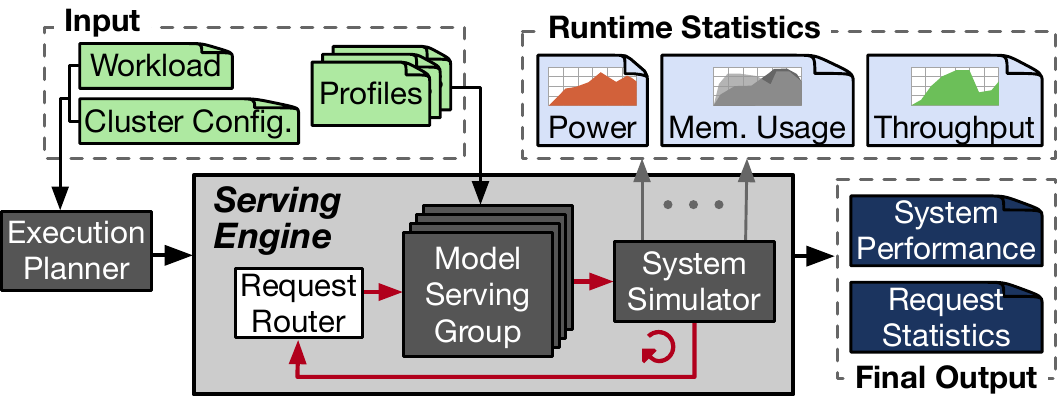}
  \caption{Overview of \simname.}
  \label{fig:overview}
\end{figure}

\subsection{Overview}
\niparagraph{Inputs.}
As shown in Fig.~\ref{fig:overview}, \simname takes three input specifications: (1) workload configuration, (2) cluster configuration, and (3) hardware performance profiles.
The workload configuration describes the LLM and request patterns, such as arrival rates and per-request execution traces.
%
The cluster configuration defines the deployment environment, including node type and count, CPU settings, memory capacity and bandwidth, and device placement. It also defines serving policies, including request routing, parallelism strategies, KV cache eviction, and compute and memory offloading decisions.

\niparagraph{Profile-based modeling.}
To enable fast and accurate evaluation at scale, \simname employs profile-based operator performance modeling.
We develop an \textit{Operator-level Profiler} built on a PyTorch/HuggingFace-based profiling API, which requires only a single device and minimal code changes, allowing users to collect device profiles without modifying model implementations.
A one-time profiling pass measures operator-level latency and power using only a single decode block per model-device pair, typically completing within 2.1 hours (e.g., Llama~3.1-70B on NVIDIA H100). 
Collected profiles can be reused across experiments without re-running hardware profiling.
%
In addition to real-hardware profiling, \simname can ingest operator-level profiles from external hardware simulators, enabling the evaluation of future accelerators such as PIM without requiring physical hardware.

\niparagraph{Serving workflow.}
Simulation begins with a one-time initialization step.
During initialization, the \textit{Execution Planner} constructs the \textit{Serving Engine} by instantiating a \textit{Model Serving Group} (MSG) per model, assigning devices and serving policies, and configuring memory, power, and system topology within the \textit{System Simulator}.
%
After this setup, execution proceeds in a runtime loop.
The \textit{Request Router} dispatches incoming requests to the appropriate MSG, which generates an execution graph for each request batch.
The System Simulator then evaluates the graph, accounting for communication, synchronization, and multi-tier memory access.
%
The loop repeats until all requests are complete, producing both online runtime statistics and final serving performance.
%

\niparagraph{Outputs.}
During execution, \simname reports system-level metrics, such as memory usage, energy consumption, and throughput.
After completion, it reports per-request serving metrics, including TTFT, TPOT, queueing delay, and end-to-end latency.
Together, these outputs offer a comprehensive view of LLM serving and enable quantitative comparisons across hardware configurations and policy choices.

\subsection{Execution Planner}
The Execution Planner takes the workload and cluster configurations and performs a one-time initialization of the Serving Engine.
It configures the Request Router to generate and route runtime requests to the appropriate MSGs.
The planner then creates an MSG for each model, assigns devices to form a customizable device pool, and installs serving policies, including parallelism strategies, compute and memory offloading, KV cache management, and memory sharing.
These flexible policies enable systematic exploration of diverse serving deployments, including heterogeneous configurations.

\begin{figure}[]
  \centering
  \includegraphics[width=0.95\linewidth]{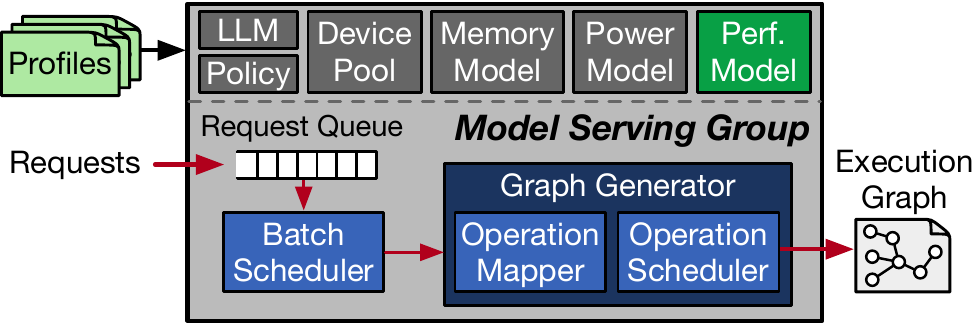}
  \caption{Model Serving Group (MSG) architecture. The MSG includes a customizable device pool enabling heterogeneous hardware composition and supports batching, operation mapping, and scheduling for LLM execution.}
  \label{fig:msg}
  \vspace{-0.1em}
\end{figure}

Each MSG is populated with operator-level performance profiles and serving policies, enabling batching, mapping, and scheduling at runtime, as illustrated in Fig.~\ref{fig:msg}.
In parallel, the planner initializes the System Simulator with cluster-level timing, network, memory, and topology configurations for system-wide performance evaluation.
After initialization, execution proceeds entirely within the Serving Engine, where the Request Router, MSGs, and the System Simulator interact iteratively until all requests complete.

\subsection{Model Serving Group}
A Model Serving Group (MSG) is a logical execution unit responsible for serving one LLM instance.
As shown in Fig.~\ref{fig:msg}, each MSG maintains a device pool containing one or more accelerators, which users may configure to reflect the desired serving deployment.
The pool serves as the execution substrate for operator mapping and may comprise a mix of accelerators and memory devices, such as GPUs, NPUs, CXL-attached devices, and PIM-equipped memory channels.
All execution decisions reflect the configured LLM, serving policies, and the simulator’s power, memory, and performance models.

\niparagraph{Request queue.}
The MSG receives requests from the router and tracks them until completion.
Each request progresses through prefill and decode while accumulating statistics such as queueing delay, TTFT, TPOT, and end-to-end latency.

\vspace{-0.1em}
\niparagraph{Batch scheduler.}
The batch scheduler selects pending requests from the queue and forms an executable batch.
It evaluates system and device memory capacity, the KV cache footprint, and the configured maximum batch size.
The memory model is consulted to determine whether the candidate batch fits within available resources, including prefix-cache residency and eviction cost.
If the batch satisfies these constraints, it is forwarded to the graph generator, where execution begins at the operation mapper.

\vspace{-0.1em}
\niparagraph{Memory model.}
The memory model governs KV cache management and device memory usage throughout inference serving.
During batch scheduling, KV cache eviction and promotion decisions are derived based on per-device memory capacity and cache residency, and the corresponding load and store overheads are injected into the execution graph.
Prefix KV cache placement across multiple memory tiers is explicitly modeled, including device memory, host memory, CXL-attached memory, and storage, while accounting for caching policies, capacity constraints, and data movement costs.
Together, these mechanisms enable accurate simulation of memory placement and movement across tiers, as well as contention for memory resources during inference serving.

\vspace{-0.1em}
\niparagraph{Power model.}
The power model tracks the instantaneous power consumption of each MSG while simultaneously accounting for the total system-level energy consumption.
For each MSG, the power model is decomposed into seven major components that account for the majority of energy consumption in modern servers: accelerators, CPUs, DRAM, interconnect links (including switches), Network Interface Cards (NICs), storage devices, and remaining system components such as the motherboard and cooling infrastructure.
%
%
Accelerator power is modeled using a three-state model comprising idle, active, and standby states to capture utilization-dependent behavior. 
DRAM and interconnect links consume energy proportional to the volume of data transferred, while the remaining components are modeled with constant power.
This design enables tracking runtime power usage across different MSG configurations and workloads, and quantifying how serving policies and system configurations influence energy consumption at both the MSG and system levels.

\niparagraph{Operation mapper.}
The operation mapper assigns operators to devices within an MSG based on the configured parallelism strategies and optional compute and memory offloading rules.
When multiple device types are available, operator placement further incorporates fine-grained operation-level offloading decisions, such as executing attention on PIM or offloading experts and KV caches to specific memory devices.
For each assignment, the simulator attaches latency and power estimates using the operator profile.
Once mapping is complete, the operator set is forwarded to the operation scheduler for dependency resolution and directed acyclic graph (DAG) construction.

\niparagraph{Operation scheduler.}
The operation scheduler constructs an execution DAG that encodes data dependencies, ordering constraints, parallelism strategies, and the required communication and memory operations.
The final output is an execution graph, which is passed to the System Simulator as input.

\subsection{System Simulator}
The System Simulator executes the operator graphs generated by each MSG and evaluates end-to-end execution at the cluster scale.
Each node in the execution graph represents an operator annotated with device type, latency, memory, and communication statistics, and power information, while edges encode data dependencies that determine scheduling and parallelism.
During runtime, the simulator models synchronization overhead, network contention, inter-device communication, and memory access latency based on cluster topology and configuration.

To support cluster-scale timing and memory evaluation, \simname builds on a modified version of ASTRA-sim~\cite{astrasim2} and Chakra~\cite{chakra}.
Those existing frameworks, originally designed for simple and repetitive training patterns, are insufficient for modeling dynamic LLM inference serving.
We therefore extend both frameworks to support heterogeneous
compute fabrics and operator-driven execution graphs that
capture the dynamics of LLM inference and support
inference-specific parallelism, such as expert parallelism.

In addition, \simname integrates a refined memory model to capture bandwidth contention, KV movement, and memory sharing with higher fidelity.
To enable such modeling, we extend the memory hierarchy to include device memory, host memory, storage, and CXL-attached memory, and explicitly model memory sharing to capture cluster-wide effects.
Furthermore, we add PIM operations beyond memory load and store primitives, enabling system-level simulation of PIM-accelerated execution. 
As a result, \simname simulates LLM inference serving by jointly modeling operator-driven execution, inference dynamics, and heterogeneous memory systems within the Serving Engine loop.

%% file: body/example.tex
\begin{figure}[]
  \centering
  \includegraphics[width=0.9\linewidth]{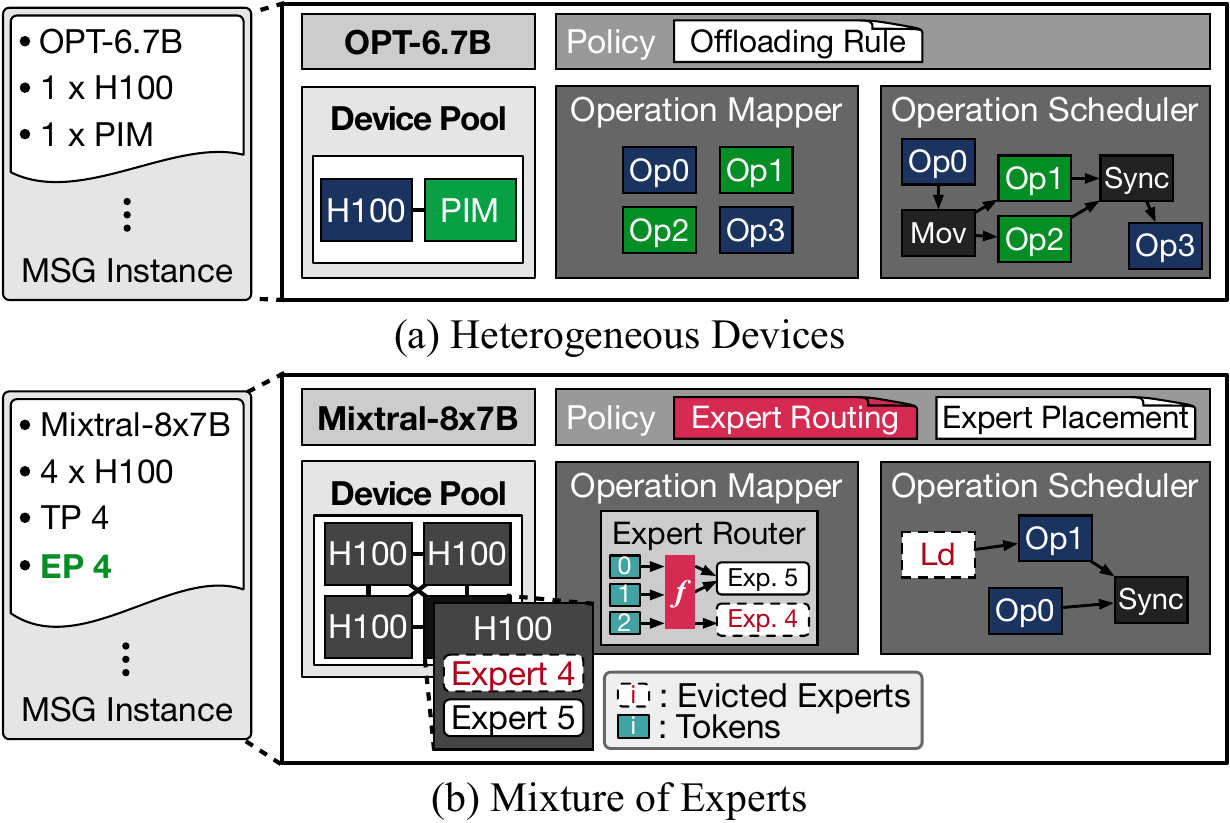}
  \vspace{-0.2em}
  \caption{Model serving group-level usage example.}
  \label{fig:msg-example}
  \vspace{-0.2em}
\end{figure}

\begin{figure}[]
  \centering
  \includegraphics[width=\linewidth]{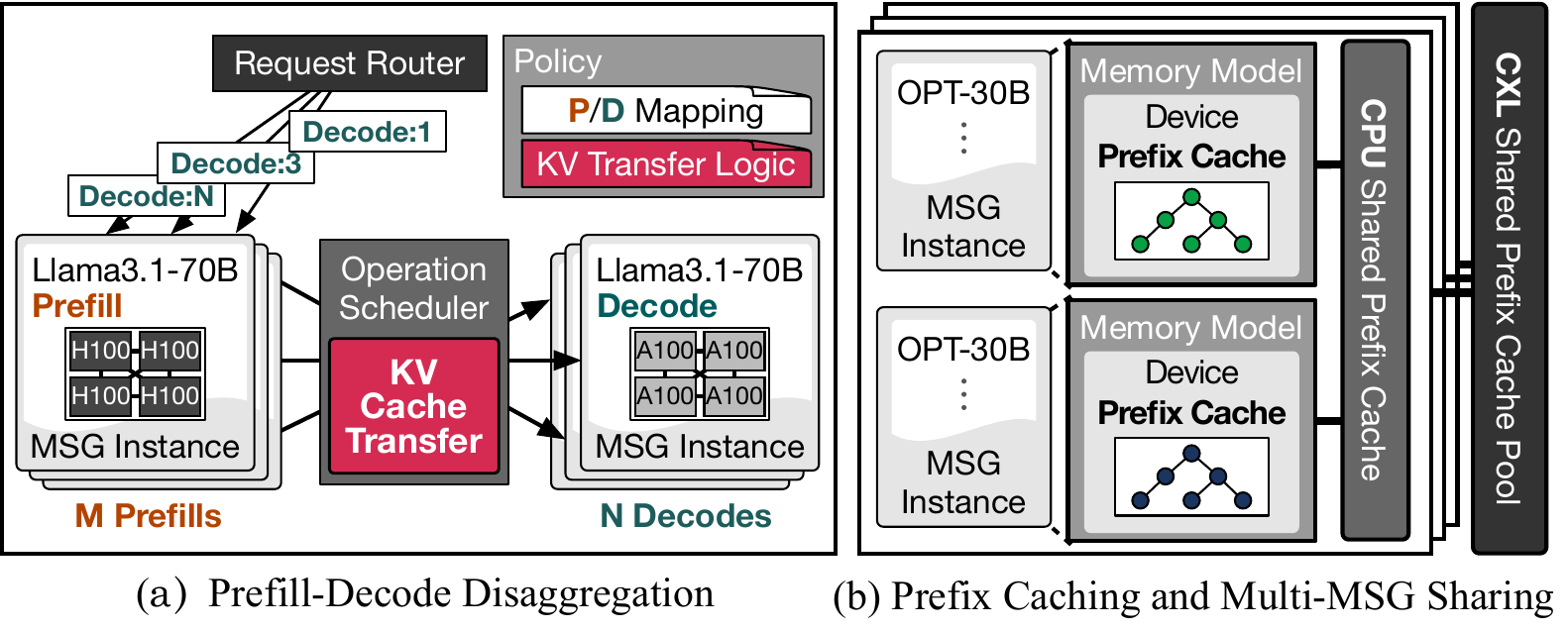}
  \vspace{-1.5em}
  \caption{System-level usage example.}
  \label{fig:sys-example}
\end{figure}

\section{Serving Techniques and Use-Case Modeling}

\subsection{Model Serving Group-Level Techniques}

\vspace{-0.1em}
\niparagraph{Heterogeneous devices and operator-granular offloading.}
\simname allows a single MSG to combine heterogeneous devices and execute an LLM with device-specialized operator placement.
During initialization, the Execution Planner constructs the MSG device pool and installs serving policies that specify how different devices are utilized.
At runtime, the MSG’s operation mapper assigns each operator to an appropriate device at operator granularity based on these policies.
As illustrated in Fig.~\ref{fig:msg-example}(a), when the device pool includes both GPUs and PIM modules and attention offloading is enabled, attention operators are mapped to PIM, while the remaining operators execute on GPUs.

The operation scheduler then incorporates the required data-movement and synchronization operations into the execution graph.
For attention-offloading, this involves transferring intermediate activations and KV caches from GPUs to PIM before execution and returning the results to GPUs afterward.
These operations are automatically inserted, executed by the System Simulator, and included in latency modeling.
Through this design, \simname supports flexible operator-granular offloading across heterogeneous devices via simple policy configuration, enabling efficient exploration of heterogeneous serving systems.

\niparagraph{MoE via expert routing, parallelism, and offloading.}
For MoE models, \simname similarly supports operator-level modeling of the full MoE execution flow within a single MSG. 
MoE behavior is controlled by serving policies installed by the Execution Planner, including expert placement across devices, expert offloading rules, and routing strategies.
Each MSG instantiates a configurable \textit{Expert Router} that emulates gating behavior and determines expert assignment on a per-token basis.
%
The router supports a variety of routing schemes, such as random selection, round-robin routing, proportional-load balancing, as well as user-defined policies.

During serving, as shown in Fig.~\ref{fig:msg-example}(b), tokens are routed to experts at each MoE layer according to the routing policy. 
Depending on the configuration, experts may reside on different devices or be temporarily evicted to host memory.
The operation scheduler automatically translates expert loading and cross-device communication into an execution graph, without requiring explicit user specification.
%
This execution graph is evaluated by the System Simulator, which accounts for latency from expert loading, token routing, and all-to-all communication in expert parallelism.
Through this mechanism, \simname enables exploration of MoE execution dynamics under diverse routing and offloading scenarios.

\subsection{System-Level Serving Techniques}

\begin{figure*}[]
  \centering
  \includegraphics[width=0.95\linewidth]{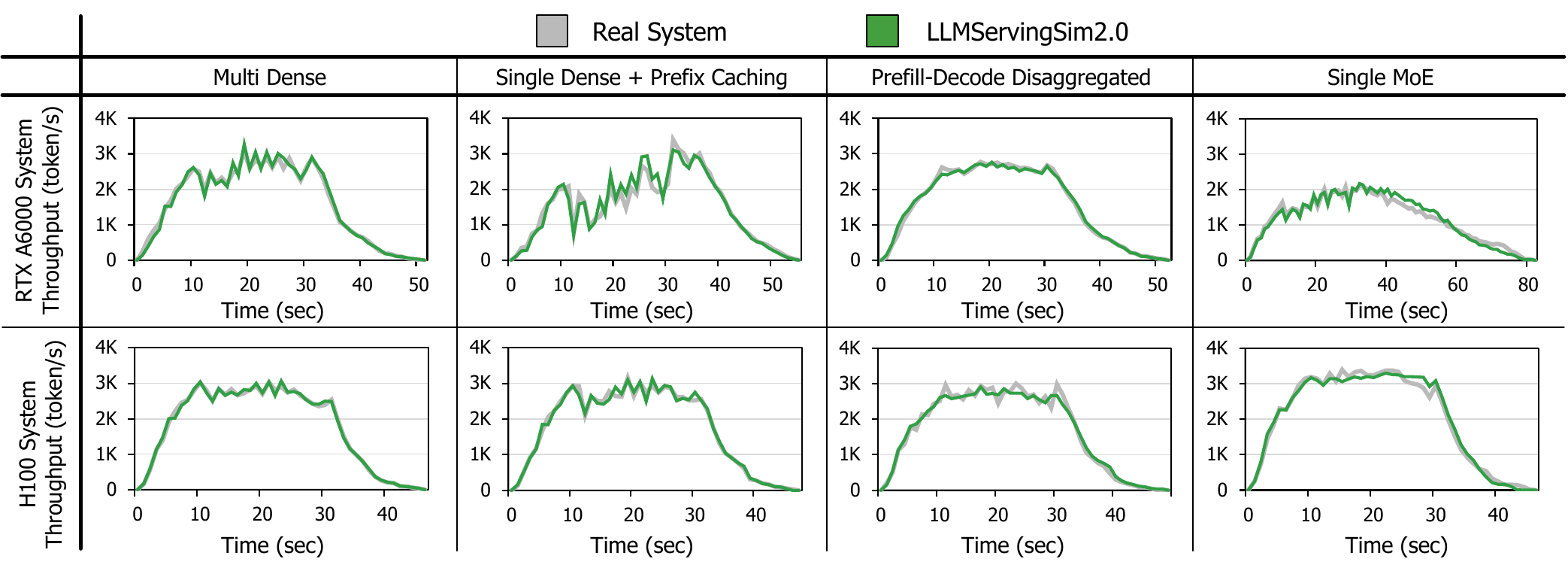}
  \caption{Comparison of throughput over time between real GPU systems (RTX A6000, H100) and \simname using the vLLM framework.}
  \label{fig:eval-main}
  \vspace{-1.5em}
\end{figure*}

\niparagraph{Prefill-decode disaggregation across serving groups.}
\simname supports PD disaggregated deployments by allowing the Execution Planner to instantiate multiple prefill and decode MSGs for the same model.
These MSGs coordinate to serve the same model but can be configured heterogeneously, with different hardware types, memory hierarchies, and interconnects. 
Based on serving policies, prefill and decode MSGs can be mapped with arbitrary M:N ratios.
As shown in Fig.~\ref{fig:sys-example}(a), at runtime, the Request Router dispatches incoming requests to a prefill MSG along with the associated decode MSG information. 
The prefill MSG operation scheduler then automatically inserts layer-wise KV cache transfer operations to the decode MSG, constructing the corresponding execution graph.
These KV movements are handled by the memory model and evaluated by the System Simulator, which computes inter-node communication latency according to the configured network topology and bandwidth.
Through this design, \simname enables exploration of PD disaggregated systems with heterogeneous hardware configurations and flexible prefill-decode mapping.

\niparagraph{Prefix caching and multi-MSG sharing.}
\simname supports prefix caching across multiple memory tiers.
Based on the configured caching policy, the Execution Planner instantiates radix-tree-based prefix caches at the appropriate tiers when constructing MSGs.
As shown in Fig.~\ref{fig:sys-example}(b), when prefix caching is limited to devices, each MSG maintains its own prefix cache within its memory model. 
When CPU host memory is used as a shared second-tier cache, MSGs on the same node retain local device-level prefix caches while additionally sharing a common CPU-resident prefix cache. 
When a CXL memory pool is enabled, all MSGs access a single globally shared prefix cache across the system.

During simulation, as requests pass through the MSG operation mapper, prefix hits reduce the effective execution latency.
If the required KV cache is not present on the device, the operation scheduler automatically inserts layer-wise KV transfer operations from the appropriate memory tier, which are executed and evaluated by the System Simulator.
With this multi-tier shared prefix cache design, \simname enables systematic exploration of diverse prefix caching policies and supports research on multi-tier caching behavior in large-scale LLM serving systems.

%% file: body/method.tex
\section{Methodology}
\label{sec:methodology}
\niparagraph{System baseline.}
We evaluate \simname on three representative serving platforms:
(1) an on-premise GPU server with four NVIDIA RTX~A6000 GPUs and an Intel Xeon Gold~6326 CPU,
(2) a cloud-based system with eight NVIDIA H100-SXM-80GB GPUs, and
(3) a TPU-v6e-1 instance on Google Cloud.
Across all platforms, we use vLLM~\cite{vllm} as the serving framework.
For model coverage, we evaluate both dense and MoE models: Llama~3.1-8B and Phi-mini MoE on the RTX~A6000 and TPU systems, and Llama~3.1-70B and Mixtral~8$\times$7B on the H100 system with a tensor parallelism degree of four.
Unless otherwise specified, all workloads used in evaluations are generated by sampling 300 requests from ShareGPT~\cite{sharegpt}, with request arrivals synthesized using a Poisson process at a rate of 10 requests per second.

\niparagraph{\simname configuration.}
To evaluate \simname, we integrate GPU and TPU backends by extracting performance models through Operator-level Profiler. 
For GPUs, we profile the target LLMs on RTX~A6000 and H100 systems and configure the simulator with matching device specifications, including memory capacity, memory bandwidth, and interconnect characteristics.
Specifically, the simulator is parameterized with 40~GB and 80~GB of device memory, 936~GB/s and 3.35~TB/s memory bandwidth, and PCIe~4.0$\times$16 and NVLink interconnects for RTX~A6000 and H100, respectively.
To demonstrate extensibility beyond GPUs, we extend the profiler to a TPU-v6e-1 instance and configure the simulator with corresponding specifications, including 32~GB of memory capacity, 1.6~TB/s memory bandwidth, and an interconnect bandwidth of 800~GB/s.
%

\niparagraph{Simulator baseline.}
We further compare the simulation accuracy and execution time of \simname against existing LLM serving system simulators, including Vidur~\cite{vidur}, APEX~\cite{apex}, TokenSim~\cite{tokensim}, and LLMServingSim~\cite{llmservingsim}.
%
%
Because these baselines do not support all serving features modeled in \simname, we limit the comparison to configurations and workloads that are executable by at least one baseline simulator.

%% file: body/eval.tex
\section{Evaluation}
\label{sec:evaluation}
\subsection{Validation with Real Serving System}

\niparagraph{Performance.}
We evaluate \simname by comparing its results against measurements from real GPU systems running vLLM, demonstrating consistent accuracy across different GPU platforms.
Fig.~\ref{fig:eval-main} shows the time-series throughput of the real system and \simname under various serving configurations, including multi-dense model serving, prefix caching, PD disaggregation, and MoE model serving.
Across both RTX~A6000 and H100 systems, \simname closely tracks the temporal throughput patterns observed in the real system.
On a time-series basis, the average throughput error is 5.14\% on RTX~A6000 and 3.29\% on H100, reflecting runtime variability from dynamic batching, request arrivals, and execution phase transitions.
This consistency holds across different serving modes and model types, despite their distinct execution and memory behaviors.
When performance metrics are aggregated over the full execution, including both throughput and latency, the average error drops to 0.99\% and 1.54\% for RTX~A6000 and H100, respectively.
These results demonstrate that \simname captures both dynamic behavior and end-to-end performance trends across diverse serving configurations and hardware platforms.

\begin{figure}[]
  \centering
  \includegraphics[width=0.97\linewidth]{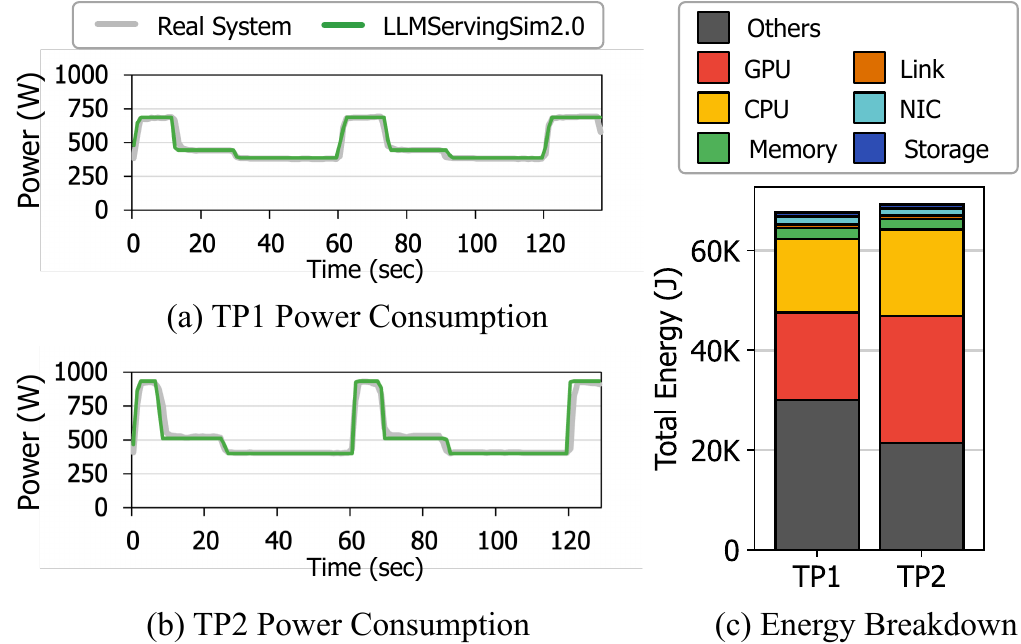}
  \caption{Comparison of power consumption and energy breakdown between the real RTX A6000 system and \simname.}
  \label{fig:eval-power}
\end{figure}

\niparagraph{Power consumption.}
%
Fig.~\ref{fig:eval-power}(a) and (b) compare the real-time power consumption of an RTX A6000 system with that predicted by \simname under tensor parallelism degrees 1 and 2. 
In this experiment, we generate each request pulse by parsing 10 requests from the ShareGPT dataset and issuing them three times at 60-second intervals. 
Idle gaps are intentionally inserted between pulses to exercise the three-state accelerator power model in \simname, which alternates among idle, active, and standby states.  

As shown in Fig.~\ref{fig:eval-power}(a) and (b), \simname closely matches the measured power consumption of the RTX A6000 system under both tensor parallelism configurations, with an average error of 1.34\% in total energy consumption.
Three distinct power pulses are observed, corresponding to transitions among active, standby, and idle states. 
%
Higher tensor parallelism activates more GPUs, resulting in higher peak power, while shorter execution time leads to narrower power peaks.
These results confirm that \simname's three-state power model and other power components accurately capture the runtime power dynamics of the real system.

Fig.~\ref{fig:eval-power}(c) presents the energy breakdown across seven power components. 
%
%
The others category is relatively large because unused GPUs in the RTX A6000 server, which contains four GPUs in total, are modeled as constant power consumers. 
Among the remaining components, accelerators dominate energy consumption, followed by CPUs and memory, consistent with a real system.
%
This accurate power modeling and fine-grained energy breakdown enable \simname to predict power consumption and evaluate energy efficiency under diverse hardware, workloads, and serving policies.
\begin{figure}[]
  \centering
  \includegraphics[width=0.97\linewidth]{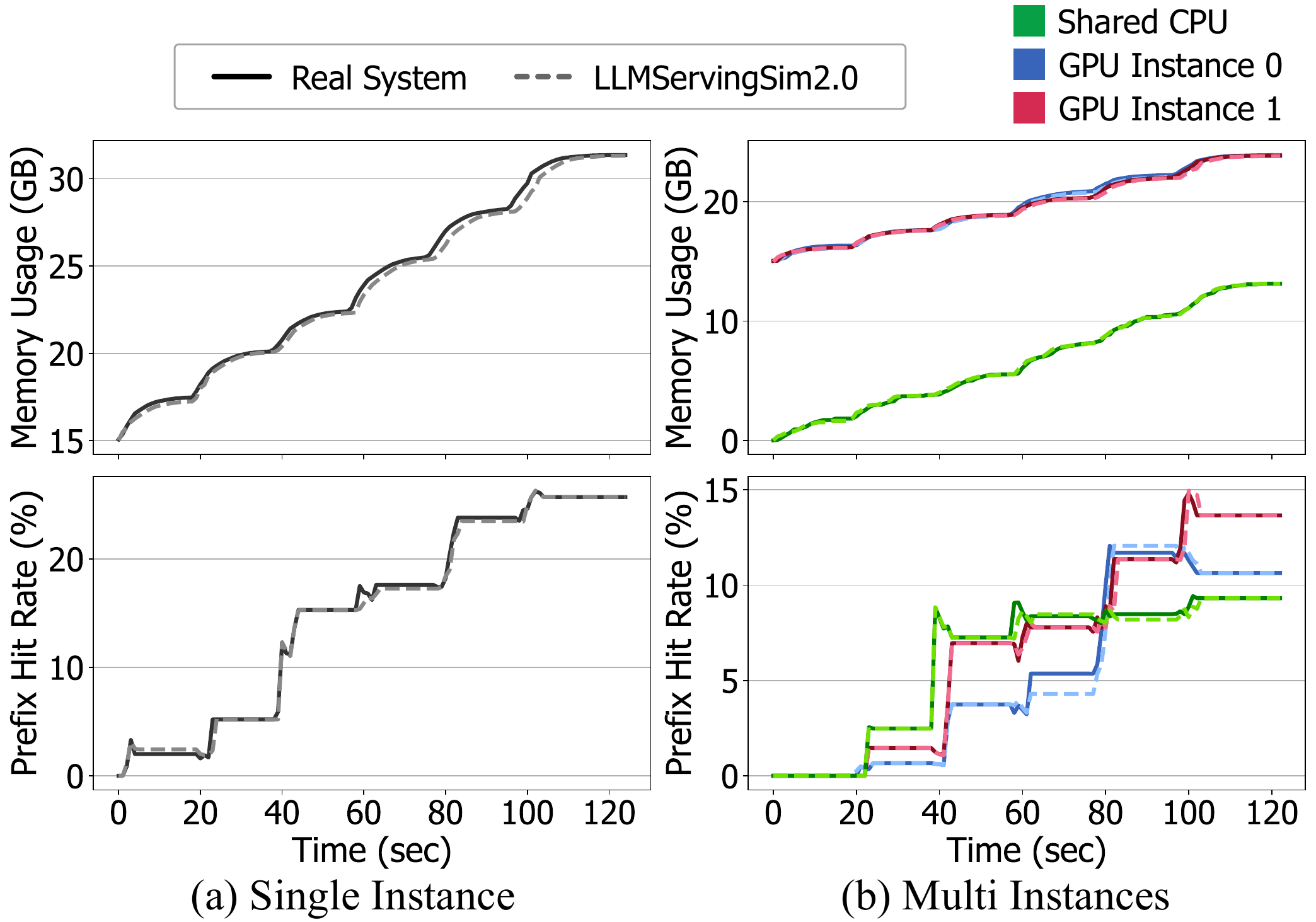}
  \caption{Comparison of memory usage and prefix hit rate between the real RTX A6000 system and \simname.}
  \label{fig:eval-memory}
\end{figure}

\begin{figure*}[]
  \centering
  \includegraphics[width=0.84\linewidth]{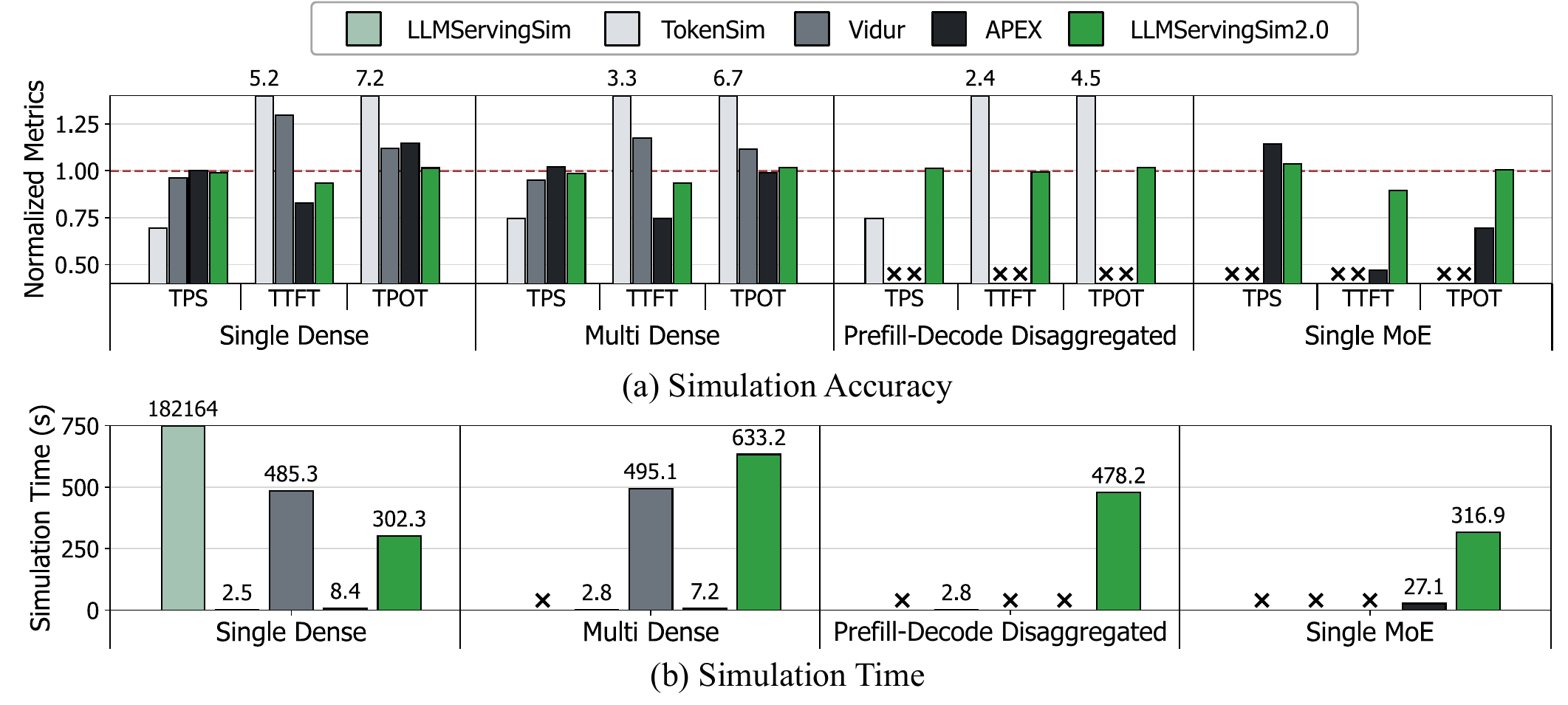}
  \vspace{-0.3em}
  \caption{Comparison with existing LLM serving simulators in terms of throughput (TPS), TTFT, TPOT, and simulation time. (a) Normalized to vLLM results.}
  \label{fig:eval-othersim}
  \vspace{-1.6em}
\end{figure*}

\niparagraph{Memory usage.}
Fig.~\ref{fig:eval-memory} compares the temporal trend of memory usage and prefix cache hit rate between RTX A6000 systems and \simname under single-instance and multi-instance serving configurations, using the same model and parallelism configuration.
For GPU-based prefix caching, we use the native vLLM prefix caching mechanism with a block size of 16.
For CPU-based prefix caching and centralized KV cache sharing, we integrate the LMCache~\cite{liu2025lmcacheefficientkvcache} framework with vLLM, configured with a block size of 256.
Workloads are sampled from ShareGPT to capture request lengths and prefix reuse patterns.
Request arrivals are synthesized with alternating bursty and idle periods to reproduce the temporal fluctuations in memory usage and prefix cache hit rates observed in real serving systems.
Fig.~\ref{fig:eval-memory}(a) shows results for a single-instance setup.
In this setting, \simname closely matches the real system in both memory usage and prefix cache hit rate, with an average error of 0.93\%, capturing the step-wise growth in device memory footprint and the corresponding increase in prefix reuse as requests accumulate.
Fig.~\ref{fig:eval-memory}(b) presents a multi-instance setting with centralized CPU-based prefix cache sharing, where two instances maintain separate device memory pools while accessing a shared CPU prefix cache pool.
In this setting, the real system exhibits dynamic memory usage across instances and a higher aggregate prefix cache hit rate enabled by cross-instance reuse.
\simname reproduces these behaviors with an average error of 0.41\%, accurately capturing per-instance memory growth and synchronized increases in prefix cache hit rate over time.
%

\subsection{Comparison with Other Simulators}

Fig.~\ref{fig:eval-othersim} compares \simname with prior LLM serving simulators in terms of simulation accuracy and time across diverse serving configurations.
We evaluate Vidur~\cite{vidur}, APEX~\cite{apex}, TokenSim~\cite{tokensim}, and the original LLMServingSim~\cite{llmservingsim}, with all performance metrics normalized to a real GPU system running vLLM.
Configurations unsupported by baseline simulators are marked as unavailable.
Fig.~\ref{fig:eval-othersim}(a) reports normalized throughput (TPS), TTFT, and TPOT.
Across single- and multi-instance dense serving configurations, \simname achieves high accuracy across all metrics, with an average error of 2.43\%.
Prior simulators tend to be accurate only for specific metrics, and show larger deviations on others.
Under more complex configurations, including PD disaggregation and MoE serving, several baselines fail to execute or rely on simplified abstractions, resulting in large errors.
%
%
In contrast, \simname remains accurate across all supported metrics, achieving an average error of 1.81\%, demonstrating its ability to jointly model serving dynamics and memory behavior under complex serving configurations.
Fig.~\ref{fig:eval-othersim}(b) reports simulation time.
Compared to lightweight simulators such as Vidur and TokenSim, \simname incurs higher simulation overhead due to its detailed modeling of serving dynamics and memory behavior.
%
%
This overhead is an inherent trade-off of LLMServingSim 2.0’s design goal: supporting heterogeneous accelerator types and hardware combinations while accurately capturing system-level interactions.
%
%
Although this leads to longer simulation time than GPU-centric baselines, it enables simulation of serving behavior on heterogeneous platforms beyond their scope.
%
%
Moreover, compared to the original LLMServingSim, \simname substantially reduces simulation time while expanding the supported design space, striking a practical balance between modeling fidelity and simulation efficiency.
%
%

\subsection{Case Study: Emerging Hardware}
\simname supports emerging hardware platforms, including TPU and PIM. 
As long as per-operator latency models are available through specification or profiling, these platforms can be incorporated into the simulation framework. 
We demonstrate this capability through case studies.

\niparagraph{TPU.}
To demonstrate \simname’s ability to integrate emerging accelerators beyond GPUs, we conduct a case study using a TPU-v6e-1 instance on Google Cloud.
We deploy a real TPU serving system using the TPU-enabled vLLM framework and extract an operator-level performance model for TPU-v6e-1 via a TPU-specific profiler.
As the current vLLM-TPU framework stably supports only single-instance dense serving, our validation is limited to this configuration.
We use Llama~3.1-8B with a tensor parallelism degree of one and reuse the same workload trace as in the GPU experiments.
The simulator is configured to match the TPU-v6e-1 memory capacity and bandwidth, enabling accurate reproduction. 

\begin{figure}[]
  \centering
  \includegraphics[width=0.83\linewidth]{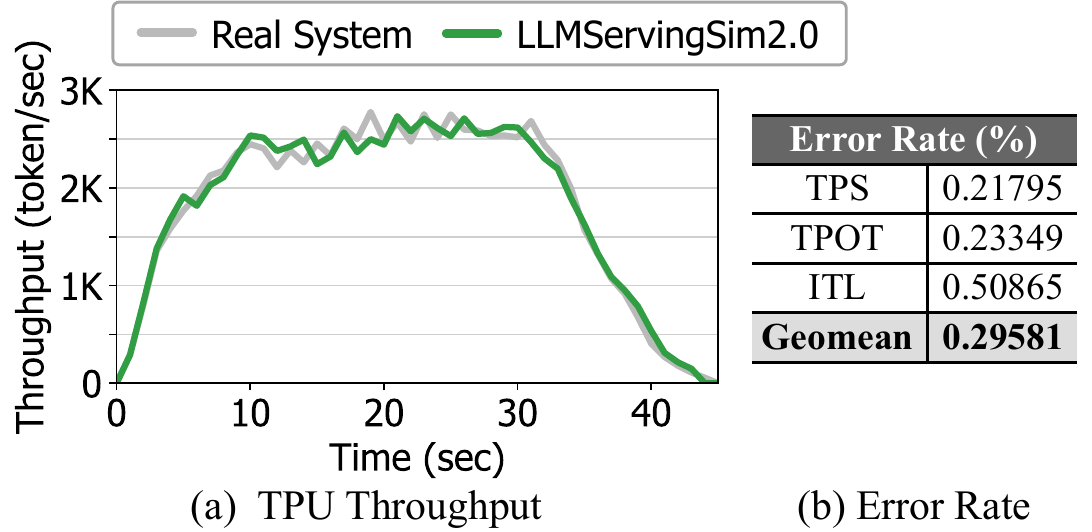}
  \caption{Comparison of throughput over time and latency metrics between real TPU system and \simname using vLLM framework.}
  \label{fig:tpu-val}
  \vspace{-0.1em}
\end{figure}

Fig.~\ref{fig:tpu-val} compares the time-series throughput of the real TPU system and \simname under the same workload.
Despite architectural differences between GPUs and TPUs, \simname closely tracks the real system, with an average per-timestep throughput error of 3.95\%.
When performance is aggregated over the full execution, the average error goes below 0.3\% across all evaluated metrics.
These results show that \simname can effectively model LLM serving behavior on non-GPU accelerators when appropriate operator-level performance models are available.
%
%
%
Moreover, \simname enables hypothetical analysis of serving techniques not yet supported by current TPU frameworks, such as PD disaggregation and prefix caching, allowing early exploration of their system-level impact.

\niparagraph{Processing-in-Memory.}
We conduct a case study comparing a GPU-only system and GPU+PIM systems, including a variant with sub-batch interleaving (SBI) proposed in NeuPIMs~\cite{neupims}.
In the GPU-only setup, Llama~3.1-8B runs on a single RTX A6000 GPU.
In the GPU+PIM setup, main memory is replaced with a PIM system composed of 256 channels, each equipped with 1~GB of HBM2 at 2000~MT/s.
%
%
%
In the PIM setup, requests are distributed across PIM channels. 
To reflect this behavior, we use a workload of 256 requests, matching the number of PIM channels, with input and output of 128 and 512 tokens.

\begin{figure}[]
  \centering
  \includegraphics[width=\linewidth]{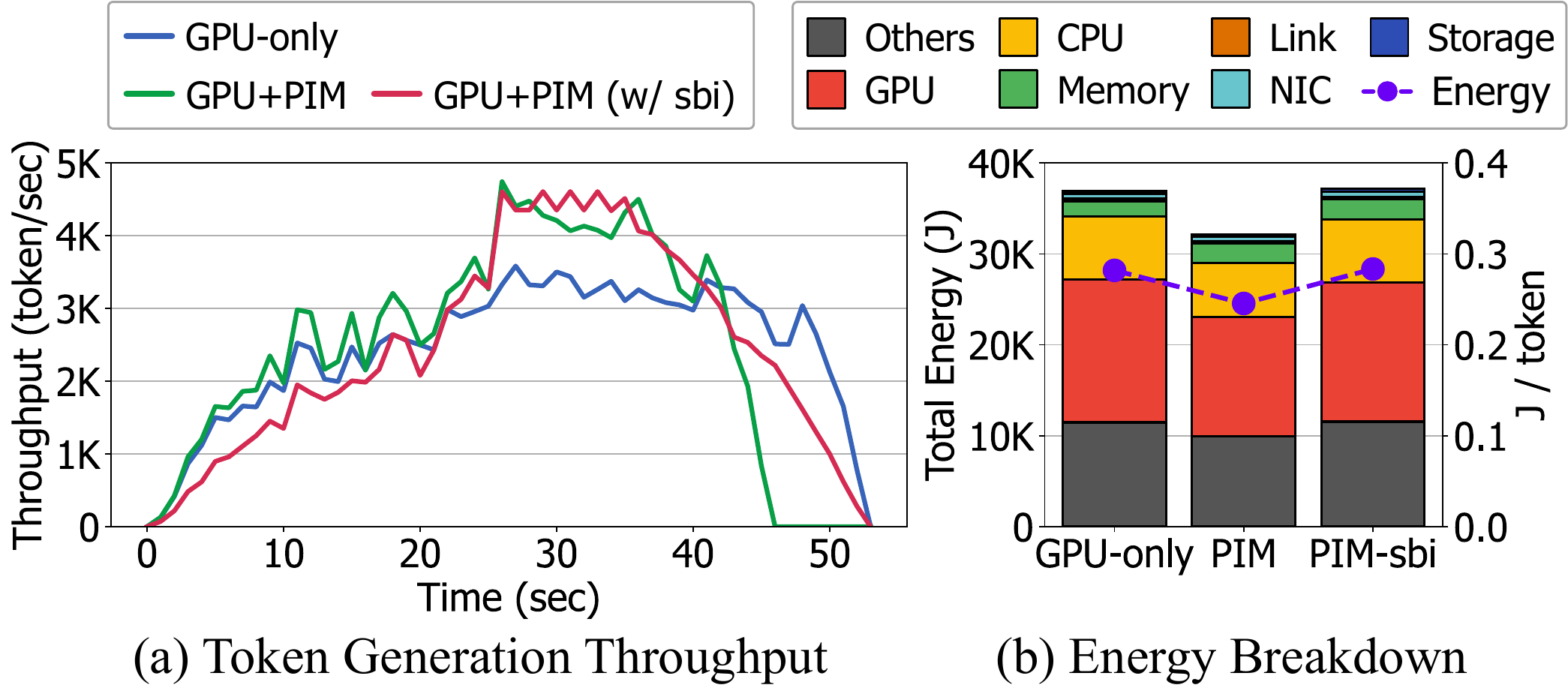}
  \caption{Performance comparison between a GPU-only system and a GPU+PIM system, including sub-batch interleaving (SBI).}
  \label{fig:eval-pim}
  \vspace{-0.2em}
\end{figure}

Fig.~\ref{fig:eval-pim}(a) compares the throughput of GPU-only and GPU+PIM systems over time.
During the prefill-dominated period (0--25\,s), performance gains are limited, as PIM primarily accelerates the memory-intensive decode phase. 
After prefill completes, GPU+PIM achieves 1.43${\times}$ higher throughput than the GPU-only system, demonstrating the effectiveness of PIM for decode workloads.
In contrast, GPU+PIM with SBI shows performance comparable to the GPU-only system. 
%
%
%
Although SBI improves hardware utilization by splitting batches, it reduces the effective batch size seen by the GPU, limiting its benefit when batch sizes are small.
As a result, SBI is effective only when very large batch sizes ($\geq 256$) are sustained, as observed during the 26--35\,s interval.

Fig.~\ref{fig:eval-pim}(b) shows the energy breakdown and J per token for the three systems.
GPU+PIM with SBI achieves performance comparable to the GPU-only system but consumes more energy due to additional PIM. 
In contrast, GPU+PIM completes requests faster and achieves the lowest energy consumption, reducing J per token by 14.8\%.
The energy breakdown shows a modest increase in memory energy with PIM, which remains small compared to GPUs and CPUs.
Overall, these results demonstrate that PIM can efficiently accelerate the decode phase of LLM serving with minimal energy overhead.

%% file: body/related.tex
\section{Related Work}
\niparagraph{Hardware-level and infrastructure-centric simulators.}
Hardware-centric simulation frameworks primarily model accelerator execution, device-level timing, and distributed communication.
LLMCompass~\cite{llmcompass} evaluates accelerator and microarchitectural trade-offs for LLM workloads, and ADOR~\cite{ador} explores hardware design choices for improving throughput.
ASTRA-sim~\cite{astrasim,astrasim2} models communication hierarchy and collective synchronization at scale, while vTrain~\cite{vtrain} analyzes parallelism strategies for training.
These systems provide valuable visibility into microarchitectural execution and communication costs, but do not model request-driven inference serving dynamics, such as batch formation, prefix reuse, KV cache behavior, or end-to-end latency under workload variation.
\simname complements this line of work by extending analysis beyond accelerator behavior to serving-time interactions and performance evolution.

\niparagraph{System-level LLM serving simulators.}
System-level simulators evaluate inference scheduling, batching, and deployment configuration.
Vidur~\cite{vidur} and APEX~\cite{apex} focus on latency/throughput prediction using operator profiling and configuration-space exploration for multi-device clusters.
TokenSim~\cite{tokensim} models dynamic arrival patterns and attention-state reuse, but relies on simplified memory abstractions that overlook bandwidth contention and KV cache movement overhead, limiting fidelity under realistic inference-serving dynamics.
LLMServingSim~\cite{llmservingsim} provides runtime statistics, including batch evolution and operator breakdowns, but supports only a single instance and does not support distributed KV cache placement or MoE-aware routing.  
In contrast, \simname supports heterogeneous device pools, profile-based execution using real or simulated hardware, and detailed modeling of KV cache movement, prefix reuse, and expert-parallel routing, allowing full-stack evaluation of serving behavior under realistic, dynamic workloads.

%% file: body/conclusion.tex
\section{Conclusion}

This paper addresses a fundamental challenge in modern LLM serving: understanding and reasoning about the complex interactions between increasingly heterogeneous hardware platforms and disaggregated serving architectures. As LLM serving systems evolve beyond homogeneous, scale-out deployments, effective system design requires tools that can capture how hardware characteristics, serving techniques, and runtime dynamics jointly shape performance, efficiency, and scalability. LLMServingSim 2.0 responds to this need by providing a unified simulation framework that makes these interactions explicit, analyzable, and extensible, enabling researchers and system designers to explore architectural trade-offs without the cost and rigidity of real deployments. By bridging hardware innovation and serving-system design within a single modeling framework, LLMServingSim 2.0 aims to serve as a practical foundation for both academic research and industrial exploration of next-generation LLM serving infrastructures.

%% file: body/acknowledgement.tex
\section*{Acknowledgments}


This work was supported by Institute of Information \& Communications Technology Planning \& Evaluation (IITP) (No.RS-2024-00396013), Electronics and Telecommunications Research Institute (ETRI) (No.RS-2025-02305453), and funded by the Korea government (MSIT). This work was also partly supported by SK hynix.

%% file: artifact/ae.tex
%
%
%
%
%






\clearpage

\appendix
\section{Artifact Appendix}

\subsection{Abstract}

\simname is a unified, runtime-driven simulator for heterogeneous and disaggregated LLM inference serving infrastructures, implemented in Python and C++. Given system configurations and request traces, it models scheduling, data movement, and contention in server-scale systems to estimate throughput and latency breakdowns. The simulator additionally supports an integrated power/energy model, MoE inference, prefill-decode disaggregation and memory disaggregation, multi-tier prefix caching, and emerging hardware platforms such as TPU, PIM, and CXL. This artifact packages the simulator and scripts/configurations needed to reproduce the key experimental results presented in the paper.

\subsection{Artifact check-list (meta-information)}


{\small
\begin{itemize}
  \item {\bf Compilation:} {gcc/g++ 11.4.0}
  \item {\bf Model:} {Llama~3.1-8B/70B, Phi-mini~MoE, Mixtral~8$\times$7B}
  \item {\bf Data set:} {ShareGPT~\cite{sharegpt}}
  \item {\bf Run-time environment:} {Ubuntu 22.04 LTS, Linux kernel 6.8.0-100-generic}
  \item {\bf Hardware:} {x86-64}
  \item {\bf Execution:} {
  \begin{lstlisting}[language=bash]
$ cd evaluation
$ ./run_all.sh
$ ./compare.sh
    \end{lstlisting}}
  \item {\bf Metrics:} {Throughput (tokens/s), latency (TTFT, TPOT, ITL, incl. p99), prefix hit rate (\%), power (W), energy (J), memory usage (MB)}
  \item {\bf Output:} {standard output, CSV files}
\item {\bf Experiments:} {Reproduce evaluation results for Figures~5-10}
  \item {\bf How much disk space required (approximately)?:} {15GB}
  \item {\bf How much time is needed to prepare workflow (approximately)?:} {5 minutes}
  \item {\bf How much time is needed to complete experiments (approximately)?:} {2 hours 30 minutes}
  \item {\bf Publicly available?:} {Yes}
  \item {\bf Code licenses (if publicly available)?:} {Creative Commons Attribution 4.0 International, MIT License}
  \item {\bf Workflow automation framework used?:} {No}
  \item {\bf Archived (provide DOI)?:} {Yes (10.5281/zenodo.18879965)}
\end{itemize}
}

\subsection{Description}

\subsubsection{How to access}
{
\begin{itemize}
    \item {Zenodo: \simname is published on Zenodo: 

    \bluetext{\url{https://doi.org/10.5281/zenodo.18879965}}}
    \item {GitHub: \simname is available on GitHub: 
    
    \bluetext{\url{https://github.com/casys-kaist/LLMServingSim}}}
\end{itemize}
}

\subsubsection{Hardware dependencies}
\simname requires an x86-64 architecture, and the simulation time may be affected by hardware differences.
For similar simulation time results, we recommend using the hardware specified in Section~\ref{sec:methodology}.

\subsubsection{Software dependencies}
\simname has been tested on Ubuntu 22.04 LTS with Python 3.10.12 and requires gcc and g++ versions 11.4.0 or higher.
Additionally, it requires the software prerequisites of ASTRA-Sim~\cite{astrasim2} and Chakra~\cite{chakra}.
To meet these software prerequisites, we provide a prebuilt Docker image. 
Reviewers can download and run the image using the provided scripts.
See Appendix~\ref{sec:installation} for details.

\subsubsection{Data sets}
We use the ShareGPT~\cite{sharegpt} dataset to generate request traces synthesized with a Poisson arrival process.
\subsubsection{Models}
We use Llama~3.1-8B/70B, Phi-mini MoE, and Mixtral~8$\times$7B for our evaluation.
Their model architectures follow decoder-only transformer variants, including both dense and MoE designs.
%
\subsection{Installation}
\label{sec:installation}
%
\label{sec:installation}
{
\begin{itemize}
\item {Clone the \simname repository.
\begin{lstlisting}[language=bash]
$ git clone --recurse-submodules https://github.com/casys-kaist/LLMServingSim.git
$ cd LLMServingSim
\end{lstlisting}
}
\item {Run Docker.
\begin{lstlisting}[language=bash]
$ ./docker.sh
\end{lstlisting}
}
\item {Build submodules.
\begin{lstlisting}[language=bash]
$ ./compile.sh
\end{lstlisting}
}
\end{itemize}
}

\subsection{Experiment workflow}
The workflow of \simname is described in Section~\ref{sec:llmservingsim2.0}, particularly in Fig.~\ref{fig:overview}.
The simulator takes (1) a workload configuration, (2) a cluster configuration, and (3) hardware performance profiles.
The hardware performance profiles are used by each MSG when generating execution graphs.
During a runtime-driven loop, the serving engine routes requests to the appropriate MSG, performs dynamic batching, and generates an execution graph under the configured serving policies, including weight offloading, expert routing, prefill-decode mapping, KV cache transfer, and prefix caching.
The System Simulator then evaluates the graph while modeling communication, contention, multi-tier memory accesses, and integrated power/energy, and returns the results to advance the next iteration.
At each iteration, \simname reports throughput, prefix hit rate, power/energy, and memory usage.
Finally, it aggregates the overall results and reports per-request latency metrics.
%

\subsection{Evaluation and expected results}
As described in Section~\ref{sec:evaluation}, our artifact evaluation consists of six experiments corresponding to Fig.~\ref{fig:eval-main} through Fig.~\ref{fig:eval-pim}. 
To reproduce them, we provide one script per figure (\path{figure_{i}.sh}) in the \path{evaluation/} folder. 
We also provide \path{run_all.sh} to execute the full evaluation pipeline at once.
{
\begin{itemize}
\item {Move to \path{evaluation/} folder.
\begin{lstlisting}[language=bash]
$ cd evaluation
\end{lstlisting}
}
\item {Run each evaluation one by one.
\begin{lstlisting}[language=bash]
$ ./figure_5.sh
$ ./figure_6.sh
...
$ ./figure_10.sh
\end{lstlisting}
}
\item {Run all evaluation at once.
\begin{lstlisting}[language=bash]
$ ./run_all.sh
\end{lstlisting}
}
\end{itemize}
}
Each script stores outputs in its corresponding folder (\path{figure_5/} to \path{figure_10/}). 
During execution, it generates \path{logs/}, \path{results/}, and \path{parsed/} subdirectories, and produces the final figure PDFs (\path{figure_{i}.pdf}).
The parsed TSV files contain metrics such as throughput, latency, simulation time, memory, or power, depending on the figure.

\begin{figure}[t]
\centering
\framebox[0.95\linewidth]{%
\begin{minipage}{0.9\linewidth}
\small
\dirtree{%
  .1 \simname.
  .2 evaluation.
  .3 figure\_5.sh~\ldots~figure\_10.sh.
  .3 run\_all.sh.
  .3 compare.sh.
  .3 figure\_5.
  .4 A6000.
  .5 config.
  .5 logs.
  .5 results.
  .5 parsed.
  .5 reference.
  .4 H100.
  .5 \vdots.
  .4 figure\_5.pdf.
  .4 figure\_5\_ref.pdf.
  .3 \vdots.
  .3 figure\_10.
  .3 artifacts.
  .4 figure\_5~\ldots~figure\_10.
  .3 README.md.
  .2 \vdots.
  .2 README.md.
}
\end{minipage}
}
\vspace{0.7em}
\caption{Directory tree of \simname~evaluation.}
\label{fig:dirtree}
\end{figure}

For verification, we provide preserved reference outputs in \path{evaluation/artifacts/} and reference figures (\path{figure_{i}_ref.pdf}) in each figure folder.
To compare numerical values of the evaluation, we provide \path{compare.sh}, which automatically compares generated parsed TSV files against the reference outputs in \path{evaluation/artifacts/figure_{i}/parsed}.
{
\begin{itemize}
\item {Compare one or more figures.
\begin{lstlisting}[language=bash]
$ ./compare.sh 5
$ ./compare.sh 6 7 8
$ ./compare.sh figure_9
\end{lstlisting}
}
\item {Compare all figures at once.
\begin{lstlisting}[language=bash]
$ ./compare.sh
\end{lstlisting}
}
\end{itemize}
}
For visual comparison, compare each generated figure PDF (\path{figure_{i}_.pdf}) with its corresponding reference PDF (\path{figure_{i}_ref.pdf}) in the same figure folder.
For more information about evaluation, please refer to \path{evaluation/README.md}.
Additional per-figure details are documented in \path{evaluation/figure_{i}/README.md}, including each figure’s objective, required configurations and datasets, exact run commands, expected generated PDF/TSV outputs, and numerical/visual validation procedures.

Fig.~\ref{fig:dirtree} illustrates the directory tree of \simname, including the per-figure evaluation scripts (\path{figure_{i}.sh}), their corresponding \path{figure_{i}} folders with generated outputs (\path{logs/}, \path{results/}, \path{parsed/}, and PDFs), the utility scripts (\path{run_all.sh}, \path{compare.sh}), the archived reference outputs in \path{artifacts/}, reference figure PDFs (\path{figure_{i}_ref.pdf}), and \path{README.md} files.

\subsection{Experiment customization}
\subsubsection{Input configurations}
\simname uses a cluster configuration JSON file as the main hardware/system input.
Cluster configuration is located in \path{cluster_config/{name}.json}, where users can customize topology and instance settings such as \path{num_nodes}, \path{link_bw}, \path{link_latency}, \path{num_instances}, \path{cpu_mem}, \path{model_name}, \path{hardware}, \path{npu_mem}, \path{npu_num}, \path{npu_group}, and \path{pd_type}.
Optional fields include \path{placement}, \path{pim_config}, \path{power}, and \path{cxl_mem}.

\subsubsection{Input dataset}
\simname uses request traces in JSONL format located in \path{dataset/{name}.jsonl}.
Each request line includes \path{input_toks}, \path{output_toks}, \path{arrival_time_ns}, and \path{input_tok_ids}.
Custom traces can be generated with \path{dataset/sharegpt_parser.py} or created manually using the same JSONL schema.

\subsubsection{Input parameters}
\label{sec:input_parameters}
\simname~\path{main.py} provides runtime options listed below. See \path{README.md} for more details.
\begin{itemize}
\item Input/output options: \\
\spath{--cluster-config}, \spath{--dataset}, \spath{--output}

\item Core options: \\
\spath{--fp}, \spath{--block-size}, \spath{--max-batch}, \\
\spath{--max-num-batched-tokens}, \spath{--num-req}

\item Routing/scheduling options: \\
\spath{--request-routing-policy} \\
\spath{--expert-routing-policy} \\
\spath{--prioritize-prefill}

\item Feature toggles: \\
\spath{--enable-prefix-caching} \\
\spath{--enable-prefix-sharing} \\
\spath{--prefix-storage} \\
\spath{--enable-local-offloading} \\
\spath{--enable-attn-offloading} \\
\spath{--enable-sub-batch-interleaving} \\
\spath{--enable-attn-prediction}

\item Run-control/logging options: \\
\spath{--gen}, \spath{--log-interval}, \spath{--log-level} \\
\spath{--network-backend}
\end{itemize}

\subsubsection{Evaluation-script customization}
For evaluation, each per-figure script (\path{figure_{i}.sh}) declares key inputs near the top of the file, including paths to cluster configurations and datasets, as well as runtime options passed to \path{main.py}.
Users can customize configuration files, workload traces, and execution settings using the input parameters listed in Appendix~\ref{sec:input_parameters}.
%
%
\subsection{Notes}
More information can be found in the \path{README.md} file of each directory.
\balance




